# PRELIMINARY CONSIDERATIONS FOR THE DESIGN OF THE INTERACTION REGION

M.E. Biagini, INFN-LNF, Frascati, Italy

*Abstract*

The design of an Interaction Region (IR) suitable to reach low values of the β functions at the Interaction Point (IP) and a high collision frequency is a rather difficult task in a short ring as DAΦNE, where an upgrade of the peak luminosity to $10^{34}$ cm$^{-2}$s$^{-1}$ is aimed [1]. In the following, general considerations on the design of such an IR are presented, together with a preliminary design; the final design will however be the result of a joint collaboration between accelerator and detector physicists.

## IR DESIGN REQUIREMENTS AND CONSTRAINTS

Design requirements and constraints depend mainly on the geometry of the detector and on the value of the β functions at the IP. Since machine and detector requirements are often conflicting the IR geometry will be a trade-off between them. In the following a list of the main items to be studied is presented, with a **D** label for detector requirements or constraints, and with an **M** for machine ones. This list is probably not exhaustive of all the IR design problems, but is just intended to give an idea of the actual difficulties we have to face.

**a)** The detector needs a free solid angle as large as possible. This means not only that the accelerator components should occupy the smallest cone around the beam pipe, but also that they should be kept as far away from the IP as reasonably possible (**D**) to allow for the installation of a vertex detector. However this is conflicting with the request to…

**b)** push the first vertical focusing quadrupole as close to the IP as possible, to minimize the beam spot size (**M**).

**c)** One of the tools to reach high peak luminosity is to increase the collision frequency. This translates, for a flat beam, into the necessity of introducing a horizontal crossing angle (**M**). The choice of its value is critical and will be discussed in detail in the following section.

**d)** The detector needs to have a large solenoidal field to reconstruct the particle kinematics (**D**). This is a very large perturbation for a low energy beam, however this has been successfully addressed and solved in DAΦNE, where a coupling as low as 0.2 % has been reached, so…

**e)** …a smart coupling correction scheme has to be designed (**M**).

**f)** The IR quadrupoles should have very small size to fit in a small cone (**D**) but…

**g)** …the closest they are to the IP, the larger are their gradients (**M**). All or some of them will need to be embedded in the detector field. Chromaticity correction can be an issue for very low values of $\beta_y^*$.

**h)** Adequate shielding from the beam backgrounds should be provided (**D+M**). Due to the low energy of the beams, in DAΦNE the main source of backgrounds are the Touschek scattered particles. Dedicated simulations to design masks and collimators are needed (**M**). The DAΦNE experience will be very useful also in this field.

**i)** To reduce $K_s$ regeneration and to improve interferometry a very thin beam pipe, with a large radius (10 cm) around the IP, is appreciated (**D**).

**j)** Ultra-vacuum is needed and the number, dimension, position and type of vacuum pumps has to be carefully studied (**M**).

**k)** The impedance budget of the vacuum pipe has to be computed taking into account all the possible sources of impedance and trying to minimize them (**M**). This work has been accomplished for the present DAΦNE IR with good results.

**l)** An "instrumented" beam pipe, with a calorimeter as close to the IP and to the beam pipe as feasible, is needed (**D**).

From all the previous points it is clear that the IR design needs to be studied from many points of view. The recipe should include geometry considerations and beam-beam simulations that are needed to set the design beam parameters (β, σ) at the IP. Of course such a complicate design has to be carried out in collaboration with the detector physicists.

## CROSSING ANGLE CHOICE

The crossing angle geometry has many advantages: it allows for a higher collision frequency (so that a larger number of bunches can collide), the beams are "naturally" separated as soon as they leave the collision point (so there is no need for dipoles close to the IP) and the beams can be sooner accommodated in two separate rings. These are the reasons why "factories", as DAΦNE, CESR and KEK-B, have chosen it. However the crossing angle geometry has also many drawbacks.

First of all, the bunch spacing is reduced so the beams travel in the same pipe with a small separation and can interact with destructive effects. This is the 'parasitic crossing' (PC) issue: a distance between beam cores of at least 10 $\sigma_x$ is required at the first PC (the most harmful) in order not to have the beam tails seeing each other, with a consequent decrease in lifetime.

Perhaps a more important effect is the reduction in luminosity and tune shifts due to the PC collision. This item, which also depends on the value of the beams separation (absolute, not in number of $\sigma_x$) will be addressed in the following section.

Moreover with a large crossing angle, highly desirable from a "geometric" point of view, synchro-betatron resonances which couple the transverse and longitudinal phase space can be excited, with a resulting increase of the beam spot size at the IP and a consequently lower luminosity. The Piwinski angle, defined as:

$$\phi = \theta\, \sigma_z/\sigma_x$$

where $\theta$ is the half crossing angle and $\sigma_x$ and $\sigma_z$ are the horizontal and longitudinal beam sizes, is a parameter used to estimate how dangerous the crossing angle can be. Up to now DAΦNE ($\phi$ =0.29) and KEK-B ($\phi$ =0.57) are the storage rings where $\phi$ has reached higher values with some loss in luminosity due to beam blow up but no destructive effects; however this parameter should in general be kept as low as possible, and it could be a limitation when trying to reach very high beam-beam tune shift values.

The parameters used in the following considerations are summarized in Table 1.

Table 1: Main parameters

| C (m) | 90. | $\varepsilon_x$ (mm mrad) | 0.2 |
|---|---|---|---|
| $\beta_x^*$ (m) | 0.5 | $\beta_y^*$ (mm) | 4. |
| N bunches | 150 | $\sigma_x^*$ (mm) | 0.3 |
| s @ 1$^{st}$ PC (m) | 0.30 | $\sigma_l$ (mm) | 3.8 |
| $\beta_x^{PC}$ (m) | 0.84 | $\beta_y^{PC}$ (m) | 18.7 |
| $\xi_x^{IP}$ head-on | 0.098 | $\xi_y^{IP}$ head-on | 0.098 |

The minimum value for the crossing angle can be set by imposing that at the first PC the two beams have a safe 20 $\sigma_x$ separation. With the chosen parameters we have:

$$\theta_{min} = \pm 15\ mrad$$

On the other hand the maximum value of $\theta$ is dictated by the detector request that the machine components occupy just a $\pm 9°$ cone around the IP (present design for the KLOE detector at DAΦNE). The value of the angle then depends on the distance L* of the first quadrupole from the IP. By choosing L* = 0.2 m (a very short distance indeed) and a small dimension quadrupole with an aperture of $\pm 10\ \sigma_x$, we get:

$$\theta_{max} = \pm 50\ mrad$$

Within these two values the Piwinski angle will range between 0.18 and 0.63. Of course by reducing $\sigma_z$ we can decrease the Piwinski angle to a safer value.

Another problem of large crossing angles comes by the off-axis trajectory in the magnetic elements, where the field quality degrades. Non-linear fields and fringing field effects have then to be carefully taken into account when modelling the beam trajectory.

The choice of the beam pipe aperture is also very important, since it affects not only the quadrupole sizes, but also the Touschek beam lifetime [2], since the scattered particles can be lost inside the IR [3]. Clearly the larger the crossing angle, the larger is the requested aperture. The resulting crossing angle will be a compromise between minimum aperture, maximum lifetime, minimum backgrounds and maximum free solid angle for the detector.

## PARASITIC CROSSINGS EFFECT

The tune shifts due to the PCs can be estimated, for Gaussian beams, by [1]:

$$\xi_x = -\frac{N\, r_e}{2\pi\gamma}\, \frac{\beta_x\,(x^2 - y^2)}{(x^2 + y^2)^2}$$

$$\xi_y = +\frac{N\, r_e}{2\pi\gamma}\, \frac{\beta_y\,(x^2 - y^2)}{(x^2 + y^2)^2}$$

where x and y are the horizontal and vertical beam separation, N is the number of particles in the opposite bunch. It can be seen from these formulae how it is just the absolute value of the separation that counts from the beam-beam point of view, and not the number of $\sigma_x$, which limits the lifetime instead.

As an example, a crossing angle of ±30 mrad has been chosen to evaluate the importance of the PC tune shift with respect to the IP one. In Fig. 1 the absolute value of the PC $\xi_y$, normalized to the IP one, is plotted as a function of the PCs position in the IR. The PC tune shift contribution has to be counted twice since each bunch experiences a PC collision on both sides of the IP.

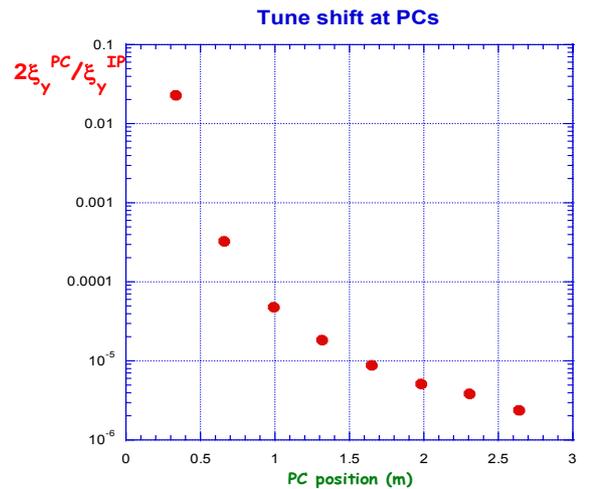

Figure 1: Normalized vertical tune shift: ratio of PC to main IP vs PC position (note the logarithmic vertical scale).

It is obvious that the main contribution (a 2% effect) comes from the first PC, where the separation is smaller (20 $\sigma_x$ for a ±30 mrad crossing angle). The second PC has a 0.4% effect only. The others are clearly negligible. The effect on the horizontal tune shift is a factor of 20 lower, due to its smaller $\beta_x$ value. It has to be confirmed by a beam-beam simulation that 2% will not have a dramatic effect on the total tune shifts. Smaller values of the crossing angle will give a more important contribution (a 10% for 15 mrad), larger will of course have much less impact (0.8% for 50 mrad), as it is shown in Fig. 2, where only the contribution from the first PC is taken into account as a function of the crossing angle.

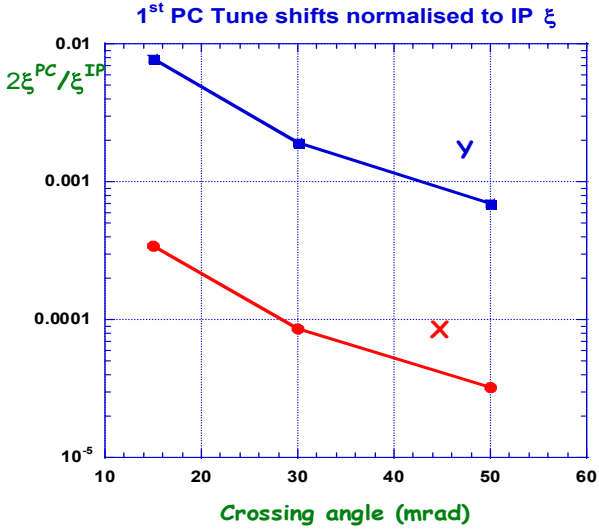

Figure 2: First PC horizontal (red dots) and vertical (blue dots) tune shift, normalised to the main one, vs crossing angle. (Note the logarithmic vertical scale).

## LUMINOSITY AND TUNE SHIFTS

The effect of a finite crossing angle on the luminosity and beam-beam tune shifts needs also to be studied with beam-beam simulations, however to give a first estimate we can use the following formulae [5], valid for any crossing angle and for $\gamma \gg tg(\theta/2)$:

$$L = \frac{N^2}{4\pi\sigma_y\sqrt{\left(\sigma_z^2 tg^2(\theta/2) + \sigma_x^2\right)}}$$

$$\xi_{xp} = \frac{r_e N}{2\pi\gamma} \frac{\beta_x}{\sqrt{\left(\sigma_z^2 tg^2(\theta/2) + \sigma_x^2\right)}\left(\sqrt{\left(\sigma_z^2 tg^2(\theta/2) + \sigma_x^2\right)} + \sigma_y\right)}$$

$$\xi_{yp} = \frac{r_e N}{2\pi\gamma} \frac{\beta_y}{\sigma_y\left(\sqrt{\left(\sigma_z^2 tg^2(\theta/2) + \sigma_x^2\right)} + \sigma_y\right)}$$

with the usual meaning of the symbols. The overall effect of the crossing angle is a reduction of both luminosity and tune shifts. In Fig. 3 the luminosity for a finite crossing angle, normalized to the head-on case is plotted as a function of the $\beta_x^*$ for three crossing angle values.

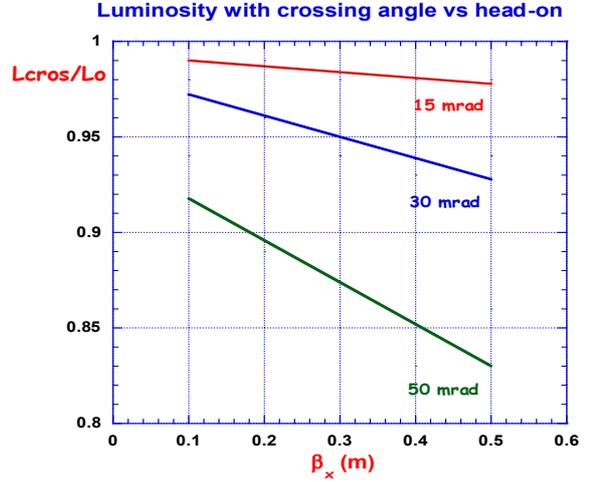

Figure 3: Luminosity with crossing angle, normalized to head-on luminosity, vs $\beta_x^*$ for different crossing angles.

As it is shown in the luminosity formula above, the effective horizontal beam size is increased by a factor ($\sigma_z$ tg ($\theta/2$)), with a consequent reduction of luminosity of the order of 14% for the design $\beta_x = 0.5$ m and an angle of ±50 mrad. In this calculation a ratio 100 between horizontal and vertical $\beta$ at the IP has been taken, while to minimize the hourglass effect a bunch length equal to the $\beta_y^*$ is taken.

The tune shifts reduction computed from the previous formulae and normalized to the design tune shifts, as a function of the horizontal $\beta$ at the IP, are plotted in Figs. 4 and 5

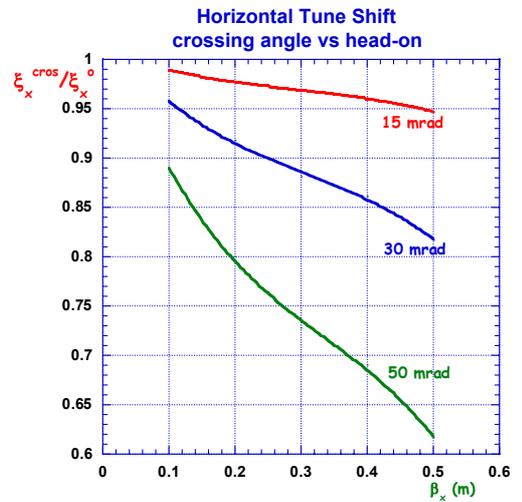

Figure 4: Beam-beam horizontal tune shift with crossing angle (normalized to design value) vs $\beta_x^*$ for different crossing angles.

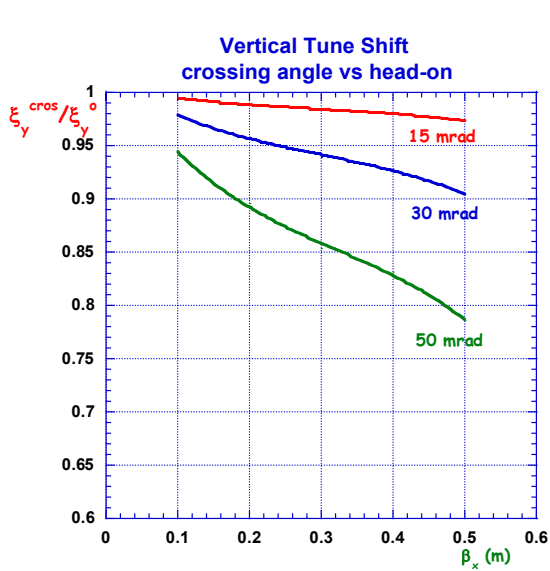

Figure 5: Beam-beam vertical tune shift with crossing angle (normalized to design value) vs $\beta_x^*$ for different crossing angles.

The vertical tune shift is reduced by the same factor as the luminosity, while the horizontal one drops faster. Due to this reduction, the beam footprint is smaller, so in principle we could still increase the luminosity by increasing the beam current.

To have a more clear picture of the problem, Fig. 6 shows the luminosity ratio (crossing/head-on) plotted versus the crossing angle, for the design values of $\beta_x^* =$ 0.5 m, $\beta_y^* =$ 4 mm and $\sigma_l =$ 3.8 mm. The reduction is about 14% for the larger angle.

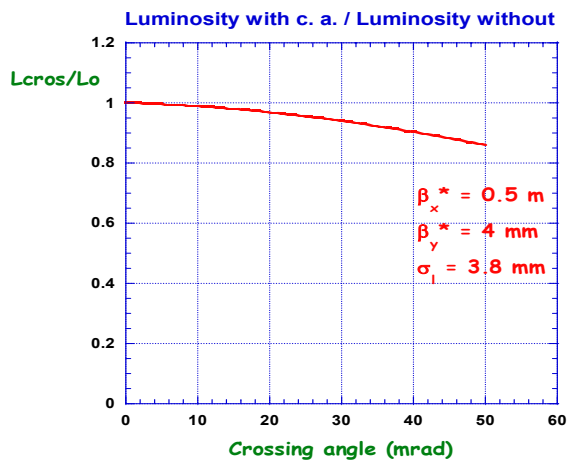

Figure 6: Luminosity reduction due to crossing angle, for the design parameters.

In Fig. 7 the tune shifts reduction for the same configuration, a factor 14% in y and 26% in x for 50 mrad, is plotted.

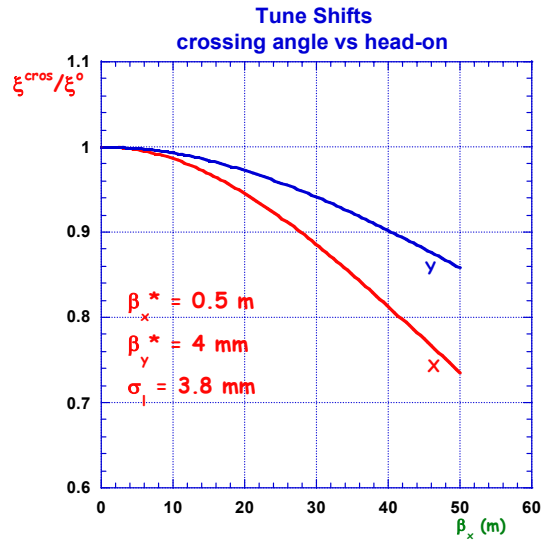

Figure 5: Beam-beam tune shift reduction vs crossing angle, for the **design parameters**.

As a conclusion on the crossing angle choice, we can say that quite a number of items enter in the game. They are:
- IR geometry
- Apertures
- PC effect on lifetime
- PC effect on beam-beam tune shifts
- Luminosity and tune shifts reduction
- Background studies

The choice will mostly depend on the trade-off between the number of colliding bunches and the reduction in tune shifts and luminosity that is reasonably tolerable. However, once a value of θ has been chosen, beam-beam and background simulations have to be performed and according to the results its value may need to be changed.

## EXAMPLE OF AN IR LAYOUT

As an example an IR design is presented in the following. A list of the main parameters that have been preliminarily chosen is in Table 2.

Table 2: IR and beam parameters

| Total length (m) | 10. | $\beta_x^*$ (m) | 0.5 |
|---|---|---|---|
| $\theta_x$ (mrad) | ± 30 | $\beta_y^*$ (mm) | 4. |
| L* (m) | 0.20 | $\varepsilon_x$ (mm mrad) | 0.2 |
| Solid angle (deg) | ± 9 | Min. quads clearance | ±10 $\sigma_x$ |

A two triplets D-F-D configuration has been chosen. The basic principle is to separate the two beam lines as soon as possible, for this reason a ±30 mrad crossing angle was chosen. In this scheme the first quadrupole (QD1, horizontally defocusing) is shared by both beams, while the second and third (QF2, QD3) should be already accommodated on separate beam lines. The beams pass off-axis in QD1 and on-axis in QF2 and QD3. With this geometry the beams separation at the entrance of QF2 is about 14 cm, at QD3 is about 60 cm, while at the IR end is about 74 cm. Due to the small space available the permanent magnets choice is mandatory, a quadrupole similar to the CESR one [6] is a possible choice. Due to the ±9° cone constraint and the small crossing angle, the quadrupole dimensions are extremely small. Moreover due to the small value of L*, the distance of QD1 from the IP, this quadrupole has to have an external radius of 3 cm, a pole radius of 1.5 cm and a thickness of 1.5 cm of pm material, very challenging values but still sufficient to provide the needed gradient, about 40 T/m. A sketch of half IR is shown in Fig. 8. The optical functions and beam half-separation, in half IR, are shown in Figs. 9 and 10 respectively.

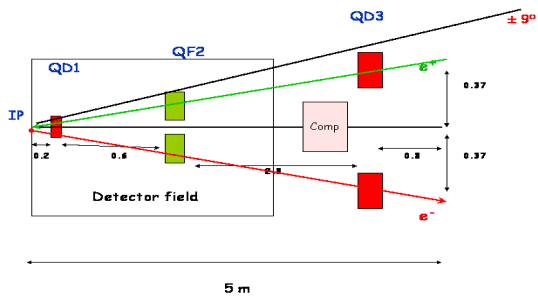

Figure 8: Sketch of half IR.

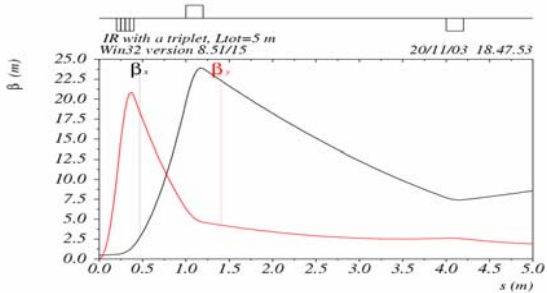

Figure 9: Half IR optical functions (x in black, y in red). The IP is at s = 0.

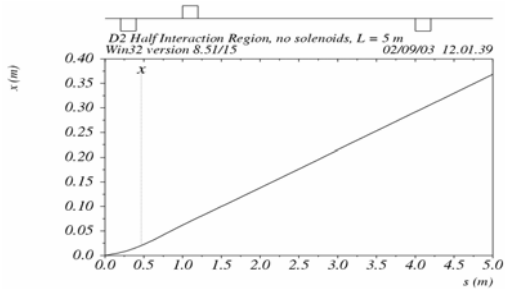

Figure 10: Beams half-separation in half IR with a ±30 mrad crossing angle. The IP is at s = 0.

In Table II the quadrupoles characteristics are summarized. The space available to separate the two beam lines between the two QF2 quadrupoles is only 4 cm in this configuration, while there is a 25 cm space available between the two QD3. This last could even be of a normal conducting type, used for matching the optical functions outside the IR. A technical design will be needed in order to check for the feasibility of such a design.

Table 2: Quadrupole parameters

|     | L (m) | G (T/m) | Pole radius (cm) | PM Thick. (cm) | Beam |
|-----|-------|---------|------------------|----------------|------|
| QD1 | 0.2 | 39. | 1.5 | 1.5 | Off axis |
| QF2 | 0.2 | 11. | 11. | 1.5 | On axis |
| QD3 | 0.2 | 3. | 15. | >1.5 | On axis |

The flexibility of the IR versus the $\beta_y^*$ value from 1.5 to 5 mm, while keeping the same quadrupole strengths, has been checked... The $\beta_y$ variation at the end of the IR ranges between 0.7 and 2. m, easily matched to the ring optical functions by adjusting the cell quadrupoles or QD3. In Fig. 11 the $\beta_y$ behaviour in half IR is plotted for different IP $\beta_y$ values. The red line corresponds to the design $\beta_y^*$ value of 4 mm.

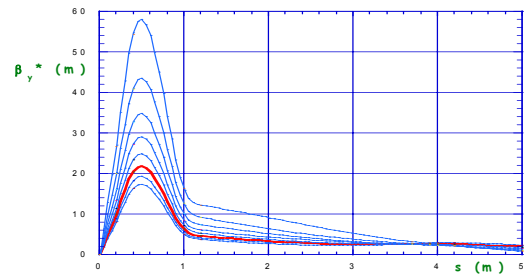

Figure 11: $\beta_y$ behaviour in half IR for different $\beta_y^*$ values.

## COUPLING CORRECTION

The coupling correction depends of course on the detector field value. At 510 MeV the beam rotation due to the solenoidal field is a large perturbation to the optics. In principle eight parameters are needed to decouple the whole IR transfer matrix, to have minimum coupling both at the IP and outside the IR. The present DAΦNE correction scheme for the KLOE detector, where all IR quadrupoles are embedded in the detector field, consists of 2 compensating solenoids plus a tilt in each quadrupole, proportional to the field integral at the quadrupole location, and of fine adjustments of the skew quadrupoles outside the IR. This scheme has proven to be very efficient, allowing for lowering the coupling to a value of about 2x10$^{-3}$.

The new IR example has two quadrupoles immersed in the detector field, this means that 2 compensating solenoids + 4 quadrupole tilts + 2 skew quadrupole in IR (can be the QD3) + skew quadrupoles outside IR are needed. Once the IR design has been finalized and the value of the detector field has been chosen, the correction scheme will be studied.

## CONCLUSIONS

A preliminary study of a simple design of an IR suitable for a high luminosity DAΦNE upgrade has been presented. A more detailed study is in progress and in particular the following items need to be addressed in the future:

- technical IR design;
- study of the pm quadrupoles;
- chromaticity correction study;
- aperture definition;
- background evaluation;
- beam pipe design;
- vacuum design;
- impedance budget;
- trapped HOM study;
- temperature control.